\pdfoutput=1

\documentclass[
	reprint,
	amsmath,
	amssymb,
	aps,
	longbibliography,
	prl,
	floatfix
]{revtex4-2}

\usepackage[utf8]{inputenc}

\usepackage{amsmath}

\usepackage{soul}
\usepackage{physics}
\usepackage{tensor}
\usepackage{slashed}
\usepackage{multirow}

\usepackage[nomessages]{fp}	

\usepackage[hyperref]{xcolor}

\definecolor{goethe-blau}{cmyk}{1.0,0.2,0.0,0.4}
\definecolor{hellgrau}{cmyk}{0.04,0.04,0.05,0.02}
\definecolor{sandgrau}{cmyk}{0.12,0.09,0.13,0.0}
\definecolor{dunkelgrau}{cmyk}{0.25,0.25,0.30,0.75}
\definecolor{emo-rot}{cmyk}{0.04,1.0,0.8,0.07}
\definecolor{purple}{cmyk}{0.08,1.0,0.3,0.36}
\definecolor{senfgelb}{cmyk}{0.01,0.25,1.0,0.05}
\definecolor{gruen}{cmyk}{0.62,0.4,0.87,0.09}
\definecolor{magenta}{cmyk}{0.08,0.86,0.12,0.12}
\definecolor{orange}{cmyk}{0.0,0.7,1.0,0.04}
\definecolor{sonnengelb}{cmyk}{0.0,0.12,0.95,0.0}
\definecolor{helles-gruen}{cmyk}{0.4,0.17,0.81,0.07}
\definecolor{lichtblau}{cmyk}{0.8,0.0,0.06,0.04}

\usepackage[
colorlinks=true,
linkcolor=goethe-blau,
breaklinks,
pdfpagelabels,
pdfstartview=FitH,
bookmarksopen=true,
bookmarksopenlevel=2,
bookmarksnumbered=true,
plainpages=false,
hypertexnames=false,
citecolor=emo-rot,
urlcolor=purple,
pdfsubject={{particle physics}}
]{hyperref}

\usepackage{bookmark}

\usepackage{upgreek}
\usepackage{cancel}

\usepackage{ wasysym }
\usepackage{orcidlink}

\usepackage{graphicx}
\usepackage{dcolumn}
\usepackage{bm}

\allowdisplaybreaks

\begin{document}

\preprint{APS/123-QED}

\title{Anomalous interactions between mesons with nonzero spin and glueballs}


\author{Francesco Giacosa \orcidlink{0000-0002-7290-9366}}
	\email{francesco.giacosa@gmail.com}
	\affiliation{
		Institute of Physics, Jan Kochanowski University,	ul. Uniwersytecka 7, P-25-406 Kielce, Poland,
	}
	\affiliation{
		Institute for Theoretical Physics, Goethe-University,
		Max-von-Laue-Str.~1, D-60438 Frankfurt am Main, Germany.
	}

\author{Shahriyar Jafarzade \orcidlink{0000-0002-4353-0760}}
	\email{shahriyar.jzade@gmail.com}
	\affiliation{
		Institute of Physics, Jan Kochanowski University,
		ul.~Uniwersytecka 7, P-25-406 Kielce, Poland.
	}

 \affiliation{Institute of Radiation Problems, Ministry of Science and Education,
 B.Vahabzade 9, AZ1143, Baku, Azerbaijan}

\affiliation{ Center for Theoretical Physics, Khazar University,
Mehseti 41 Street, Baku, AZ1096, Azerbaijan}

\author{Robert D. Pisarski \,\orcidlink{0000-0002-7862-4759}}
	\email{pisarski@bnl.gov }
	
	\affiliation{
		Department of Physics, Brookhaven National Laboratory, Upton, NY 11973.
	}

\date{\today}
\begin{abstract}
Topologically nontrivial fluctuations control the anomalous interactions for the $\eta$ and $\eta ’$
pseudoscalar mesons. We consider the anomalous interactions for mesons with higher spin, the heterochiral nonets with $J^{P C} = 1^{+ -}$ and $2^{- +}$. 
Under the approximation of a dilute gas of instantons,  the mixing angle between non-strange and strange mesons decreases strongly as $J$ increases, and oscillates in sign. Anomalous interactions also open up new, rare decay channels. For glueballs, anomalous interactions indicate that the $X(2600)$ state is primarily gluonic.
\end{abstract}
  
\maketitle

\textit{Introduction:}
Quantum Chromodynamics (QCD) is close to the chiral limit, where the up, down and strange quarks ($u$, $d$, and $s$)
are very light.  Consequently, when the global chiral symmetry is spontaneously broken in the vacuum,
from $SU(3)_L \times SU(3)_R \times U_A(1) \rightarrow SU(3)_V$, 
nine pseudo-Goldstone bosons should appear.
Instead, there are only eight: the usual octet of pions, kaons, and the $\eta$ meson, while the $\eta'$ is
much heavier than expected.

This occurs because the axial $U_A(1)$ symmetry of the classical theory is broken by quantum effects, through 
the anomaly of Adler, Bell, and Jackiw \cite{Adler:1969gk,Bell:1969ts}.  This splits the singlet $\eta'$ meson from the
octet mesons, and gives it a mass through fluctuations which are topologically nontrivial
\cite{Belavin:1975fg,tHooft:1976rip,tHooft1976,Gross:1979ur,Nair:2022yqi}.
The most familiar example are instantons: classical solutions of the gluon field equations
in Euclidean spacetime \cite{Belavin:1975fg}, whose effects
can be computed semiclassically \cite{tHooft:1976rip,tHooft1976}.
While instantons dominate at high temperature, in vacuum
truly quantum fluctuations also contribute \cite{Nair:2022yqi}.

While anomalous interactions are especially dramatic for the pseudoscalar multiplet, it is natural to ask how the axial anomaly affects other mesons, such as conventional mesons with higher spin, or unconventional ones, such as glueballs.  
As both mesons with nonzero spin and glueballs are massive, the effects of the axial anomaly are more subtle, affecting the mass splittings, mixing, and decays of some fields in these multiplets.

In Ref. \cite{Giacosa:2017pos}, mesons are divided into ``heterochiral'' and ``homochiral''.  In the chirally symmetric phase, heterochiral mesons are a mixture of
a left-handed anti-quark and a right-handed quark (or vice versa), as for the pseudo-Goldstone bosons.  Homochiral
mesons are formed just from a left (or right) handed anti-quark and a quark.  These begin with the vector mesons, $J^{PC} = 1^{--}$: the 
$\rho_\mu (770)$, $\omega_\mu (782)$, and $\phi_\mu (1020)$ mesons.

The anomalous interactions between heterochiral and homochiral mesons are very different.  Heterochiral
mesons have anomalous interactions with no derivatives, which directly affect their mass spectrum, and with few derivatives, which affect their decays.  In contrast, homochiral mesons only have anomalous interactions
with many derivatives, through the Wess-Zumino-Witten term \cite{Gomm:1984at}.

In this Letter we construct the anomalous interactions for the underlying quark operators, and their counterparts
for heterochiral mesons and for the pseudoscalar glueball, in a dilute gas of instantons (DGI).  
After reviewing the well known case of $J=0$, the
extension to heterochiral mesons with spin $J=1$ and $J=2$,
and then with a glueball, is direct.
Because of the axial anomaly, massless quarks have exact zero modes, so that
instanton contributions to anomalous operators can be computed by saturating
these operators with these zero modes \cite{tHooft:1976rip,tHooft1976}.
The only change with nonzero spin is that the vertices which tie the zero modes differs.

At the outset we acknowledge that the topological structure of the vacuum is surely
more complicated than a dilute gas of instantons \cite{Nair:2022yqi}.  
Nevertheless, the anomalous operators which we compute in this work are novel,
and we expect a DGI to give a first estimate of their magnitude. 
Indeed, a recent analysis of the chiral phase transition near
the chiral limit suggests that a DGI may well
{\it under}estimate the effects of topologically 
nontrivial fluctuations \cite{Pisarski:2023xyz}.

The present analysis is meant to motivate further study from  
numerical simulations on the lattice, and especially from
experiment.  Thus we concentrate on phenomenology, notably
the splitting and mixing between mesons in a given multiplet, and on new decay channels which open up for mesons and glueballs.

\textit{Heterochiral multiplets:}
Mesons are classified according to their quantum numbers under spin, parity, and charge conjugation, $J^{PC}$. The total spin $J = L+S$ is the sum of angular momentum $L$ and the spin $S$, with $P=(-1)^{L+1}$ and $C=(-1)^{L+S}$. 
With massless quarks, classically left and right handed quarks are invariant the symmetry group of ${\cal G}_{\rm cl} = SU_L(3) \times SU_R(3)\times U_{A}(1)$:
\begin{equation}
    q_{L,R} \longrightarrow {\rm e}^{\mp\mathrm{i}\alpha/ 2} \, U_{L,R}\; q_{L,R} \; .
    \label{trans_phi}
\end{equation}
Here $q_{L,R} = \mathbb{P}_{L,R} \, q$, where $\mathbb{P}_{L,R}
=\frac{1}{2}(1\mp \gamma_5)$.
$U_L$ and $U_R$ are flavor rotations in $SU_L(3)$ and $SU_R(3)$, respectively, while $\exp(\mp i \alpha/2)$ is a rotation for axial $U_A(1)$.
This transformation relates nonets with the same spin and opposite parity. 

A heterochiral meson with spin zero is proportional to the quark operator
$\overline{q}_L q_R$; those of higher spin are given just by
inserting powers of the covariant color derivative,
$\overleftrightarrow{D}_\mu$, between the quark fields.
 For $J=0$, $1$, and $2$, these are $\Phi$, $\Phi_{\mu}$ and $\Phi_{\mu\nu}$, as shown in Table  (\ref{tab:chiral-transformations}). Because $\overleftrightarrow{D}_\mu$ only acts upon color and not flavor, these mesons all transform identically under chiral rotations \cite{Giacosa:2017pos}.
\begin{table}[h]
   	\centering		\renewcommand{\arraystretch}{2.}
		\begin{tabular}[c]{|c|c|}
			\hline
			Chiral Nonet &  ${\cal G}_{\rm cl}$ \\
			\hline 
			$\Phi =\overline{q}_L q_R/M_0^2$ &  $e^{\mathrm{i}\,\alpha}U_L^\dagger \Phi U_R$ \\
		\hline
		$\Phi_{\mu} =\mathrm{i}\overline{q}_L \overleftrightarrow{D}_{\mu}q_R/M_1^3$  & $e^{\mathrm{i}\alpha}U_L^\dagger \Phi_{\mu} U_R$ \\
		\hline		$\Phi_{\mu\nu} =\overline{q}_L( g_{\mu\nu}\overleftrightarrow{D}^2/4-\overleftrightarrow{D}_{\mu}\overleftrightarrow{D}_{\nu})q_R/M_2^4$ &$e^{\mathrm{i}\alpha}U_L^\dagger\Phi_{\mu\nu} U_R$ \\
			\hline
		\end{tabular}
		\caption{Heterochiral fields for 
multiplets with spin zero, one, and two.} 
\label{tab:chiral-transformations}
					\end{table}

Typically bosonic fields in an effective Lagrangian have dimensions of mass.  To ensure this it is necessary to introduce the dimensionful constants $M_0$, $M_1$, and $M_2$ for $J=0$, $1$, and $2$ in Table  (\ref{tab:chiral-transformations}).  Since the spin is increased by inserting more powers of $\overleftrightarrow{D}_\mu$, the power of $M$ increases with $J$, $\sim 1/M_J^{J+2}$. A major concern in the phenomenological analysis below is the relative magnitude of these mass scales.  

The unbroken symmetry group of the quantum theory is not
${\cal G}_{\rm cl}$, but 
${\cal G}_{\rm qu} = SU_L(3) \times SU_R(3)$ \cite{Giacosa:2017pos}.  Each $SU(3)$ contains
the element $U = \exp(2 \pi i/3)$, which generates an
abelian $Z(3)$ subgroup.  Anomalous interactions violate
$U_A(1)$, but are invariant under this $Z(3)$.
For spin zero, this begins with the cubic invariant, $\sim \det(\Phi)$, in Eq. (\ref{detphi}).   
 The  anomalous interactions for fields with higher spin, Eqs. (\ref{anom_j1}), (\ref{anom_j2}), (\ref{anom_j012}), and (\ref{anom_j022}), generalize this term. 
 
We begin by reviewing the experimental evidence for heterochiral multiplets.

\textit{ (i) Heterochiral mesons with $J=0$:}
Besides the usual pions and kaons, there are the flavor eigenstates,  $\eta_{N}\equiv\sqrt{1/2}\,(\bar{u}u+\bar{d}d)$  and $\eta_{S}\equiv\bar{s}s$. 
 Because of the axial anomaly, Eq. (\ref{detphi}), these mix to form the physical  
$\eta$ and $\eta'$ states: 
\begin{equation}
\left(
\begin{array}
[c]{c}%
\eta(547)\\
\eta^{\prime}(958)
\end{array}
\right)  =\left(
\begin{array}
[c]{cc}%
\cos\beta_0 & \sin\beta_0\\
-\sin\beta_0 & \cos\beta_0%
\end{array}
\right)  \left(
\begin{array}
[c]{c}%
\eta_{N}\\
\eta_{S}%
\end{array}
\right)  \text{ ,}
\label{mixingrot}
\end{equation}
  The mixing angle, 
$\beta_0=-43.4^{\circ}$ \cite{Kloe2}, is large and negative.  This demonstrates that the axial anomaly ensures that the physical states are closer to the octet and singlet configurations, respectively \cite{Feldmann:1998vh,tHooft:1986ooh}. In all they
form a pseudoscalar nonet, $P_{ij}=\frac{1}{2}\Bar{q_j}\mathrm{i}\gamma^5 q_i$.  The assignment for the scalar mesons, with $J^{PC}=0^{++}$, is still under debate \cite{Pelaez:2015qba,Sarantsev:2021ein,Rodas:2021tyb,Binosi:2022ydc,Klempt:2022qjf,Klempt:2021nuf,Klempt:2021wpg,Guo:2022xqu}
\footnote{Evidence is mounting toward the identification with the Particle Data Group (PDG) resonances  $S=\{a_{0}(1450),$ $K_{0}^{\ast}(1430),$ $f_{0}(1370),$
$f_{0}(1500)/f_{0}(1710)$\}, with elements $S_{ij}=\frac{1}{2}\Bar{q_j}q_i$.}. 
In all, $\Phi=S+\mathrm{i} P$, Table (\ref{tab:chiral-transformations}).

\textit{(ii) Heterochiral mesons with $J=1$:}
The pseudovector mesons with $J^{PC}=1^{+-}$ corresponds to 
$P_{\mu}$=$\big\{b_1(1235), \overline{K}_{1B}\equiv K_1(1270)/K_1(1400)$  \footnote{The kaonic pseudovector $\overline{K}_{1B}$  is included in both physical states $K_1(1270)$ and $K_1(1400)$ \cite{Divotgey:2013jba,Hatanaka:2008gu} which leads  $m_{K_{1B}}=1.31$ GeV according to \cite{Divotgey:2013jba}. 
In the PDG \cite{Workman:2022ynf}, the isoscalar strange-member of the orbitally excited vector meson has been recently identified with resonance $\phi(2170)$, see however also the discussion of Ref. \cite{Piotrowska:2017rgt}.}, $h_1(1170)$, $h_1(1415)\big\}$ \cite{Workman:2022ynf}. The mixing angle between $h_1(1170)$ and $h_1(1415)$ takes the same expression as in Eq. (\ref{mixingrot}),
$\beta_1$.
The value of $\beta_1$ is not yet known, and is discussed below. Their chiral partners with $J^{PC}=1^{--}$ are the orbitally excited vector mesons $S_{\mu}=\{\rho(1700), K^{\ast}(1680), \omega(1650),\phi(2170)\}$. The full multiplet is $\Phi_{\mu}=S_{\mu}+\mathrm{i}P_{\mu}$, Table (\ref{tab:chiral-transformations}).

\textit{(iii) Heterochiral mesons with $J=2$:}
The pseudotensor mesons $J^{PC}=2^{-+}$ listed in the PDG \cite{Workman:2022ynf}, denoted as $P_{\mu\nu}=\big\{\pi_2(1670), K_{2P}\equiv K_2(1770)/K_2(1820)$ \footnote{We identify $K_{2P}$  with $K_2(1770)$ based on the phenomenological analyses performed in Ref.\cite{Koenigstein:2015asa}. }, $\eta_2(1645), \eta_2(1875)\big\}$, are members of the heterochiral nonet with spin $2$. The isoscalar mixing analogous to Eq. (\ref{mixingrot}) via the angle $\beta_2$, is under debate, but according to the phenomenological studies of Refs. \cite{Koenigstein:2016tjw,Shastry:2021asu}, it might be large. 
The chiral partners of the pseudotensor mesons are expected to be the orbitally excited tensor mesons $S_{\mu\nu}$ with $J^{PC}=2^{++}$
\footnote{
These states are still unsettled. According to the relativistic quark model \cite{isgur1985}, orbitally excited tensor mesons have to be larger than $2$ GeV.
Indeed, there are few isoscalar tensor mesons above $2$ GeV, such as the  e.g. $f_2(2010)$, $f_2(2150)$, $f_2(2300)$, and $f_2(2340)$ \cite{Workman:2022ynf} that could be part of this nonet.}.
The full multiplet is $\Phi_{\mu \nu}=S_{\mu \nu}+\mathrm{i}P_{\mu \nu}$, Table (\ref{tab:chiral-transformations}).

\textit{Instanton induced interactions: } 
It is well known that instantons generate
the interaction \cite{tHooft:1976rip,tHooft1976,tHooft:1986ooh}
\begin{align}
&\mathcal{L}_{\text{eff}}^{J=0}=-\frac{k_0}{3!}\Big(\mathrm{det}\big(\overline{q}_L q_R\big)+\mathrm{det}\big(\overline{q}_R q_L\big)\Big)\, .
\label{qk_spin0}
\end{align}
Anticipating later results, we introduce the $J$-dependent coupling
\begin{align}    k_J= (8\pi^2)^{3}\int_0^{\Lambda^{-1}_{\overline{\text{MS}}}}d\rho\; n(\rho)\; \rho^{9 + 2 J}\, .
\label{eq:kJ}
\end{align}
This is a weighted average over the  instanton density $n(\rho)$,
which for three massless quarks and  three colors is given by  \cite{Pisarski:1980md,Gross:1980br,Boccaletti:2020mxu}: 
\begin{align}
    n(\rho)=\exp\Big(-\frac{8\pi^2}{g^2(\rho \Lambda_{\overline{\text{MS}}})}-7.07534\Big)\frac{1}{\pi^2\rho^5}\Big(\frac{16\pi^2} {g^2(\rho\Lambda_{\overline{\text{MS}}})} \Big)^{6}
   \, .\label{inst-den}
\end{align}
The expression for the running coupling constant $g(\rho \Lambda_{\overline{\text{MS}}} )$ is given
in Eq. (13) ( of Supplementary Material) to two loop order, 
while the instanton density  
$n(\rho \Lambda_{\overline{\text{MS}}} )$ is illustrated in Fig. (\ref{fig:density}).  See the Supplementary Material (SM)
for further details.  Taking 
the renormalization mass scale
$\Lambda_{\overline{\text{MS}}}=300$~MeV \cite{Workman:2022ynf}, 
for $J=0$ we obtain $k_0 \approx 2.57\cdot 10^6 \,\text{GeV}^{-5}$
\footnote{As  $\Lambda_{\overline{\text{MS}}}$ is the only mass scale in our computation, for different values of $\Lambda_{\overline{\text{MS}}}$ the $k_J$ change according to their mass dimension.}. 

Assuming that the effective bosonic field $\Phi$
is proportional to the quark bilinear \cite{tHooft:1986ooh,tHooft:1999cta,Rosenzweig:1981cu},
\begin{equation}    \mathcal{L}_{\text{eff}}^{J=0}=-a_0\Big(\text{det}\,\Phi+\text{det}\,\Phi^\dagger \Big)\;.
\label{detphi}
\end{equation}
The bosonic coupling $a_0$ depends on $k_0$
and the constant $M_0$ in Table \ref{tab:chiral-transformations},
$ a_0=k_0M_0^6/48 >0 $.

The mixing angle of Eq. (\ref{mixingrot}) is then \footnote{The numerator of the mixing angle $\beta_0$ is proportional to the renormalization constants $Z_{\eta_N}Z_{\eta_S}\approx 2.6$, which is found from the fit of experimental data to the extended Linear Sigma Model in Ref. \cite{Parganlija:2012fy}. }
\begin{align}
    \beta_0=\frac{1}{2}\tan^{-1}\Big(\frac{- \, 2.6\sqrt{2}a_0\phi_N}{(m_{\eta^\prime}^2-m_{\eta}^2)\cos{2\beta_0}}\Big) <0 ,
    \label{mixPS}
\end{align}
where the chiral condensate of non-strange quarks  $\phi_N$ can be expressed in terms of the pion decay constant $\phi_N=\bra{0}\eta_N\ket{0}\simeq1.7 f_{\pi}\approx 160$~MeV. A dilute gas of instantons gives negative $\beta_0$, in agreement with phenomenology.
Imposing the phenomenological value 
$\beta_0 = -43.6^\circ$
and using the parameters of Refs. \cite{Parganlija:2012fy,Kovacs:2016juc},  
\begin{equation}
    a_0 = 1.3 ~{\rm GeV} \; ; \; M_0 = 170~{\rm MeV} \; ,
    \label{a0}
    \end{equation}
so that the value of $M_0$ is close to that for $\phi_N$.

\begin{figure}[h]
        \centering       \includegraphics[scale=0.5]{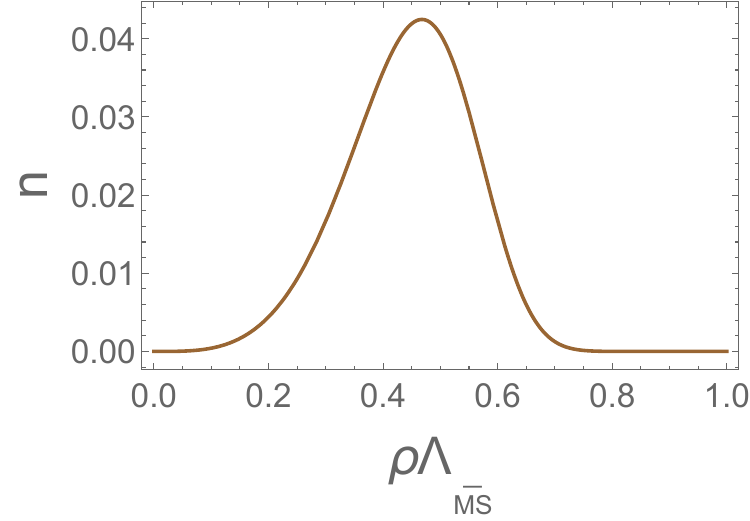}\\
        \caption{The density of instantons for $N_c = N_f = 3$.}
        \label{fig:density}
\end{figure}

\begin{figure}[h]
        \centering       \includegraphics[scale=0.65]{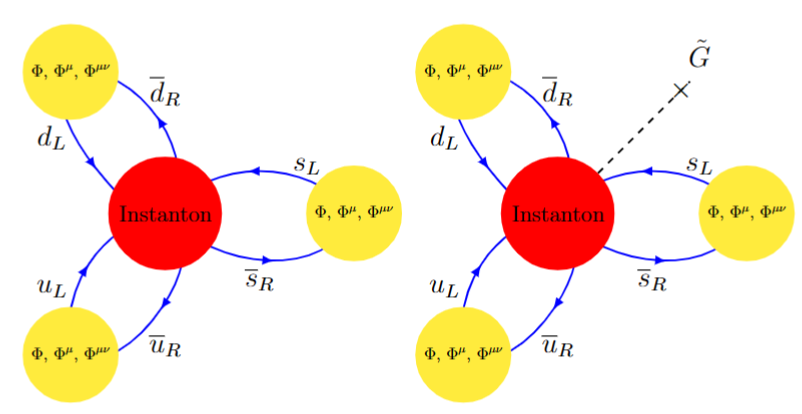}\\
        \caption{Anomalous processes induced by instantons: to left, cubic couplings between the heterochiral-type, $\Phi$'s, and to the right, their coupling to a glueball field, $\widetilde{G}$.}
        \label{fig:int}
   \end{figure}
   
The generic anomalous interaction for three flavors is
illustrated in the left part of
Fig. (\ref{fig:int}).  The only change
with higher spin is that as $J$ increases, powers of 
$\overleftrightarrow{D}_\mu$ are inserted between the zero modes.
This is responsible for the factor of $\rho^{2J}$ in the anomalous
interactions, $k_J$ in Eq. (\ref{eq:kJ}).

For spin one, the simplest anomalous interaction is quadratic
in $\Phi_\mu$ and linear in $\Phi$:
\begin{align}\nonumber \mathcal{L}_{\text{eff}}^{J=1}
&=-\frac{k_1}{3!}\Big(\epsilon\Big[\big(\overline{q}_L q_R\big)\big(\overline{q}_L\overleftrightarrow{D}_\mu q_R\big)^2\Big]
+R\leftrightarrow L \Big)\\
&=a_1 \Big( \epsilon[\Phi\, \Phi_\mu \, \Phi^\mu]
+\text{c.c.}\Big)
\, ,
\label{anom_j1}
    \end{align}
where we introduce the symbol 
\footnote{
In Ref. \cite{Giacosa:2017pos} 
instead of $\epsilon[A BC]$ we used the notation ${\rm tr}(A \times B \cdot C)$.  We prefer the former because as a product of two $\epsilon$ symbols, it is manifestly symmetric:  $\epsilon[ABC] = \epsilon[BAC]$, 
{\it etc.} 
}
\begin{equation}
\epsilon[ABC]=\epsilon^{ijk}\epsilon^{i'j'k'}A_{ii'}B_{jj'}C_{kk'}/3! \; ,
\end{equation}
with $i,j,k$ and $i',j',k'$ are $SU_L(3)$ and $SU_R(3)$ indices.
Since $\epsilon[AAA] = \det A$,
$\epsilon[ABC]$ represents a type of generalized determinant between dissimilar matrices\footnote{
We note that the operators considered here are just those of the lowest mass dimension \cite{Parganlija:2012fy,Kovacs:2013xca,Grahl:2013pba,Grahl:2014fna,Eser:2015pka,Pisarski:2019upw,Pisarski:2023xyz}.  For spin zero, 
anomalous operators with higher mass dimensions include those from
$Q = \pm 1$, such as $\sim {\rm tr}(\Phi^\dagger \Phi) \det \Phi$, and 
those from $Q = \pm 2$, such as $\sim (\det \Phi)^2$.
Obviously, there are also anomalous operators with higher mass dimensions for spin one and above.  
}.  Given the transformation properties 
of $\Phi$ and $\Phi_\mu$ in Table
(\ref{tab:chiral-transformations}), Eqs. (\ref{detphi}) and (\ref{anom_j1}) are manifestly invariant under $SU_L(3)\times SU_R(3)$. 
Similarly, as the product of three heterochiral fields, these terms are not invariant under $U_A(1)$, but $Z(3)$.  
These anomalous interactions were first obtained
in Ref. \cite{Giacosa:2017pos} entirely from
considerations of symmetry.  In this paper we now compute their
magnitude, as well as anomalous glueball interactions, in a DGI.

To relate the $k_J$ to physical processes, we need the values for
the constants of proportionality $M_J$ between quark and mesonic operators. For spin one, we find
$k_1 = 9.91\cdot10^6\,\text{GeV}^{-7}$, 
which for $M_1 = M_0$ gives
\begin{equation}
    a_1=-\frac{k_1M_1^6M_0^2}{48}\approx -0.14 ~\text{GeV}\ < 0,\,.
    \label{a1}
\end{equation}

\begin{figure}[h]
       \centering       \includegraphics[scale=0.6]{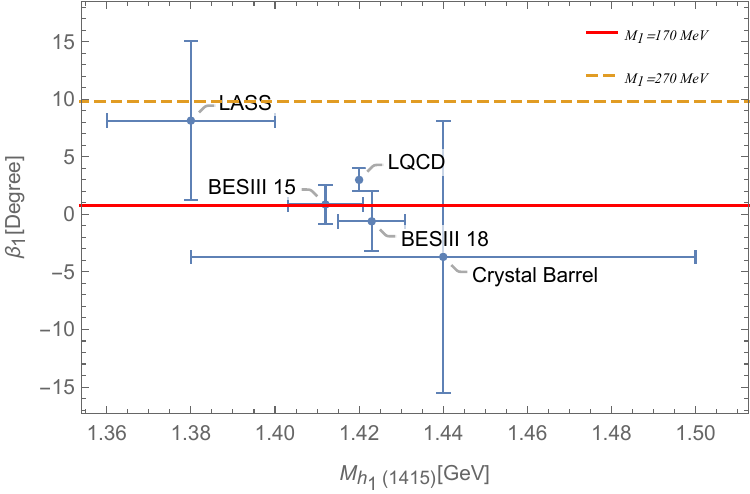}\\
      \caption{$\beta_1$ in a DGI compared to the experiment 
\cite{Aston:1987ak,CrystalBarrel:1997kda,BESIII:2015vfb,BESIII:2018ede}
      and the lattice (LQCD) \cite{Dudek:2011tt}, for $M_1 = M_0 = 170$~MeV and $M_1 = 270$~MeV. }
        \label{fig:mixing}
 \end{figure}

 The corresponding mixing angle is approximately:
 \begin{align} 
 \beta_1&\simeq\frac{1}{2}\tan^{-1}\Big(\frac{-\sqrt{2}a_1\phi_N/3}{2(m_{K_{1B}}^2-m_{b_1}^2)-\sqrt{2}a_1\phi_S/6}\Big)\ >0.
 \label{mixPV}
 \end{align}
For a DGI this mixing angle is positive.
Using the value for a strange quark condensate $\phi_S=\bra{0}\eta_S\ket{0}\approx 130$~MeV, and
assuming $M_1 = M_0 = 170$~MeV, we obtain a small value of 
$\beta_1\simeq 0.75^{\circ}$;
for a larger value of $M_1 = 270$~MeV,
the mixing angle increases to $\beta_1 \simeq 10^{\circ}$.
As illustrated in 
Fig. (\ref{fig:mixing})
\footnote{The x-axis in Fig. (\ref{fig:mixing})
      is the mass of $h_1(1415)$ meson, which represents 
      experimental uncertainty in the mixing of $P_{\mu}$ states 
      using the 
      Gell-Mann Okubo relation \cite{Gell-Mann:1962yej,Okubo:1961jc}}, experimental results favor a positive value
\cite{Aston:1987ak,CrystalBarrel:1997kda,BESIII:2015vfb,BESIII:2018ede},
as do numerical simulations on the lattice \cite{Dudek:2011tt}.

The anomalous interactions in Eq. \eqref{anom_j1} also 
open up new decay modes.  For example,
$\Gamma(\rho(1700) \rightarrow h_1(1415) \, \pi) = 0.027(M_1/M_0)^6\, \text{MeV} $, which if $M_1=M_0$
is rather small.  Other anomalous decays are discussed
in the Supplementary Material.
Measuring such processes can be used to fix the value of $M_1$.

An interaction term with one $J=0$ meson and two $J=2$ 
heterochiral mesons is
\begin{align}   \nonumber
\mathcal{L}_{\text{eff}}^{J=2} =
  & -\frac{k_2}{3!} \Big(\epsilon\Big[\big(\overline{q}_L q_R\big)\Big(\overline{q}_L(\overleftrightarrow{D}_{\mu}\overleftrightarrow{D}_{\nu}-g_{\mu\nu}\overleftrightarrow{D}^2/4) q_R\Big)^2\Big]
 \\& +R\leftrightarrow L\Big) =-a_2  \Big(\epsilon\Big[\Phi \, \Phi_{\mu\nu} \, \Phi^{\mu\nu}\Big]+\text{c.c.}\Big)
    \, .
\label{anom_j2}    \end{align}
We find $k_2 = 4.05\cdot10^7\,\text{GeV}^{-9}$, so 
when $M_2=M_0$,
\begin{equation}
    a_2=\frac{k_2M_2^8M_0^2}{48}\approx 0.017 \,\text{GeV} >0\;.
    \label{a2}
\end{equation}

The mixing angle for the pseudotensor multiplet is negative \footnote{The result for $\beta_2$ is derived by considering similar mass relations for spin-2 mesons, Eqs. (3.1)-(3.4) of Ref. \cite{Jafarzade:2022uqo}, adding the anomalous coupling in Eq. (\ref{anom_j2}).},
\begin{align}  
\beta_2&\simeq\frac{1}{2}\tan^{-1}\Big(\frac{-\sqrt{2} a_2\phi_N/3}{2(m_{K_{2P}}^2-m_{\pi_2}^2)-\sqrt{2}a_2\phi_S/6}\Big)\, <0.
\label{mixPT}
\end{align}
Assuming that $M_2= M_0$, the DGI gives a small mixing angle, $\beta_2\approx -0.05^{\circ}$.
This agrees with lattice QCD \cite{Dudek:2013yja}, but not with the large value of $\beta_2\simeq -42^{\circ}$  extracted in Ref. \cite{Koenigstein:2016tjw}  from the decay rates. 
To fit such a large mixing angle requires $M_2 = 2.4 \, M_0$.

We see that anomalous terms generate mixings between the octet
and singlet for {\it all} (pseudo-)heterochiral mesons.
These mixing angles do decrease strongly with $J$, for two reasons.
First, comparing the values in
Eqs. (\ref{a0}), (\ref{a1}), and (\ref{a2}),
each $a_J$ decreases by about $\approx 1/10$ as $J$ increases by one
({\it assuming} that $M_0=M_1=M_2$).
This is because anomalous coupling $k_J$ in Eq. 
(\ref{eq:kJ}) involves $\rho^{2J}$, and a DGI peaks at small $\rho \Lambda_{\overline{MS}}\sim 0.5$, Fig. (\ref{fig:density}).  Second, 
$\tan{\beta_J}$ is proportional to the inverse of the mass squared of the mesons, Eqs. (\ref{mixPS}), (\ref{mixPV}) and (\ref{mixPT}).  For $J=0$,
the $\eta$ and $\eta'$ are pseudo-Goldstone bosons, and so 
{\it much} lighter than ordinary mesons, with $J=1$ and $2$.
The former {\it may} be an artifact of a dilute gas of instantons; the latter is not.  Further, that the {\it sign}
of $\beta_J$ flips as $J$ changes is dynamical, and does
not follow just from the chiral symmetry.
This is a nontrivial test of our model, and appears to agree
with experiment.

Besides mixing terms, there are also
anomalous terms which involve derivatives
of the spin zero field $\Phi$, and so exclusively affect decays.
For example, a term which couples heterochiral mesons with
$J=0$, $1$, and $2$ is
\begin{align}
   \mathcal{L}_{b_2}= -b_2 \Big(\epsilon\Big[ \big(\partial_\mu \Phi\big) \Phi_{ \nu}\,\Phi^{\mu\nu}\Big]+\text{c.c.}\Big)\; .
  \label{anom_j012}
\end{align}
In a DGI $|b_2|=k_2 M_0^2 M_1^3 M_2^4/48$; with
$M_0=M_1=M_2$, $|b_2|\approx 0.099$.

An anomalous interaction coupling two heterochiral $J=0$ mesons to a
$J=2$ meson is
\begin{align}
    \mathcal{L}_{c_2}= - c_2 \Big(\epsilon\Big[\big(\partial_\mu\Phi\big)\big(\partial_\nu \Phi\big)\Phi^{\mu\nu}\Big]+\text{c.c.}\Big)\,.
\label{anom_j022}   
\end{align}
For a DGI,  $|c_2|=k_2 M_0^2 M_2^4/48$, with $|c_2|= 0.474\,\text{GeV}^{-1}\,$ when $M_2 = M_0$.  

Again, numerous anomalous decay channels open up.
For example, $\Gamma(\eta_2(1870) \rightarrow \rho(1700) \, \pi)= 1.5\cdot 10^{-6}  \; M_1^3 M_2^4/M_0^7$ MeV. Measuring such processes will {\it significantly}
constrain the values of $M_1$ and $M_2$, and test the consistency of our approach.

Besides anomalous mesonic interactions, those involving
glueballs follow immediately, and are 
illustrated in the right part of Fig. \ref{fig:int}.
An anomalous interaction between a pseudoscalar glueball
and heterochiral mesons is given by the term
\begin{align}\label{eq:gl}
   \mathcal{L}_{c_g}= -\mathrm{i}c_{g}\Tilde{G}_0\Big(\text{det}\,\Phi-\text{det}\,\Phi^\dagger \Big)\,.
\end{align}
In a DGI $c_g\approx 11$. Then, by using 
Ref. \cite{Eshraim:2012jv}, we obtain  $\Gamma(\Tilde{G}_0\rightarrow K\bar{K}\pi)\approx 0.24$ GeV and $\Gamma(\Tilde{G}_0\rightarrow \pi \pi \eta^\prime) \approx
0.05$ GeV.  In contrast to the anomalous decays between
heterochiral mesons, these are {\it large} values.
Notably, the BESIII collaboration has recently seen a pseudoscalar resonance, denoted as $X(2600)$, in the $\pi \pi \eta^\prime$ channel \cite{BESIIICollaboration:2022kwh}.
Our results support the interpretation of this resonance as mostly gluonic, with a decay enhanced by the chiral
anomaly \footnote{We note that the mass found in lattice QCD \cite{Athenodorou:2020ani} agrees with the 
identification of this state as a glueball.}. 

Further anomalous decays involving heterochiral
mesons with higher spin follow directly, and
include interactions such as  $\Tilde{G}_0\Big( \epsilon[\Phi\, \Phi_\mu \, \Phi^\mu] -\text{c.c.}\Big)$.


We conclude by noting that there are {\it many}
other anomalous interactions which can be 
computed with our techniques.  These include 
baryon decays \cite{Olbrich:2017fsd}, tetraquarks \cite{Fariborz:2007ai}, glueballs and hybrid states \cite{Eshraim:2020ucw,Shastry:2022mhk}, 
and the $H$ di-baryon \cite{Gongyo:2017fjb,Green:2021qol}.  
In summary, the effects of the axial anomaly
merely begin with the $\eta$ and the $\eta '$
mesons, but most certainly do not end there.

\bigskip
\begin{acknowledgments}
The authors thank Fabian Rennecke and Adrian K{\"o}nigstein for useful discussions. 
R.D.P. was supported by the U.S.
Department of Energy under contract DE-SC0012704,
and by the Alexander von Humboldt Foundation.
F.G.\ acknowledges support from the \textit{Polish National Science Centre} (NCN) through the \textit{OPUS} project 2019/33/B/ST2/00613.
S.J.\ acknowledges financial support through the project \textit{Development Accelerator of the Jan Kochanowski University of Kielce}, co-financed by the \textit{European Union} under the \textit{European Social Fund}, with no. POWR.03.05. 00-00-Z212 / 18 and is thankful to the Nuclear Theory divison of the Brookhaven National Labaratory for warm hospitality during his visit in which the current project was initiated.

\end{acknowledgments}

\bibliography{main}

\begin{widetext}

\section{SUPPLEMENTAL MATERIAL}

\section{Details on the instanton solution}
Instantons are self-dual solutions of the classical Euclidean equation of motion of QCD. The representation of a singular gauge instanton solution involves three parameters: the instanton location $z$, instanton size $\rho$, and gauge group orientation $U$:
\begin{align}
A^{a}_{\mu}=-2U\,\overline{\eta}^{a\mu\nu}\,U^\dagger\frac{\rho^2(x-z)^\nu}{g(x-z)^2((x-z)^2+\rho^2)}\,,
    \end{align}
    where the t'Hooft symbol is defined as
\begin{align}
\overline{\eta}^{a\mu\nu}:=\delta^{\alpha\nu}\delta^{\mu4}-\delta^{\alpha\mu}\delta^{\nu4}+\varepsilon^{a\mu\nu}\,,
\end{align}
 and for the anti-self-dual solution $\overline{\eta}\rightarrow \eta$
\begin{align}
\eta^{a\mu\nu}=:-\delta^{\alpha\nu}\delta^{\mu4}+\delta^{\alpha\mu}\delta^{\nu4}+\varepsilon^{a\mu\nu}\,.
\end{align}
The corresponding fermionic zero-modes of the Dirac operator are 
\begin{align}\nonumber
\big[\Psi_f(x_f,z,\rho,U)\big]_i^a=\frac{\sqrt{2}\,}{\pi}\frac{U^a_{\;k}\,\rho}{\big((x_f-z)^2+\rho^2\big)^{\frac{3}{2}}}
\frac{\big[\gamma^\mu(x_f-z)_\mu\big]_{ij}}{|x_f-z|}\varphi_R^{jk}
\end{align}
where $a\in\{1,\cdots, N_c\}$ is the color index, $i,j,k\in\{1,2\}$ are the spinor indices, the anti-symmetric tensor $\varepsilon$ is included within $\varphi_R=\frac{1}{\sqrt{2}\,}\begin{pmatrix}
  0\\\varepsilon
\end{pmatrix}$ and three Pauli $\vec{\sigma}$ matrices are used to represent the gamma matrices in the chiral representation, 
\begin{align}
    \gamma^{\mu}=\begin{pmatrix}
        0&i\sigma^\mu\\
        -i\overline{\sigma}^\mu&0
    \end{pmatrix}\,,\qquad \overline{\sigma}^{\mu}:=(\vec{\sigma},i\openone_2)^{\mu} \,,\qquad\sigma^{\mu}:=(\vec{\sigma},-i\openone_2)^{\mu}\, ,
\end{align}
To obtain the quark zero-modes in the case of an anti-instanton, we perform the substitution $\varphi_R\rightarrow\varphi_L=\frac{1}{\sqrt{2}\,}\begin{pmatrix}
  \varepsilon\\ 0
\end{pmatrix}$.

The quark zero-mode is approximated using a free fermionic propagator at a large distance from the instanton centre $z$:
\begin{align}\label{limit}
\Psi_f(x_f,z,\rho,U)\xrightarrow{x_f\longrightarrow\infty}  (2\sqrt{2}\,\pi \rho) \Delta(x_f-z)U\varphi_R\,,
\end{align}
where
\begin{align}    \Delta(x):=\frac{\gamma_\mu x^\mu}{2\pi^2(x^2)^2}\,.
\end{align}

The zero-mode determinant for the diagonal source $\mathcal{J}_{fg}(x_f):=\mathcal{J}(x_f)\delta_{fg}$ reads
\begin{align}\label{0-det}
    \mathrm{det}_{0}(\mathcal{J})=\underset{fg}{\mathrm{det}}\Big[\int d^4x_f\,d^4z\,\Psi_{f}^\dagger(x_{f},z,\rho,U)
    \mathcal{J}_{fg}(x_f)\Psi_{g}(x_{g},z,\rho,U)\Big]\,,
\end{align}
which reduces to the $N_f$-point Green's functions in which two fermion lines connect the source to the origin, as shown in Fig. \ref{fig:int}.

\section{From instantons to effective Lagrangians}
An ansatz for the effective Lagrangian in terms of the quark zero-modes and the isospin half Dirac spinor $\omega_j$ given in Ref. \cite{tHooft1976} is written as: 
\begin{align}
\mathcal{L}_{\text{eff}}= \frac{1}{N_f}\prod_{f=1}^{N_f}\Bigg[\Big(\Psi^\dagger(x_f,z,\rho,U)J(x_f)\Psi(x_f,z,\rho,U)\Big)+\Big(\Psi^\dagger_f(x_f,z,\rho,U)\,\omega_j \overline{\omega}_j\,\Psi_f(x_f,z,\rho,U)\Big)\Bigg]
\end{align}
which leads to the following form of the amplitude with the source $\mathcal{J}(x_f)$ :
\begin{align}
\prod_{f=1}^{N_f}\Big(\overline{\omega}_j\Delta(x_f-z)\mathcal{J}(x_f)\Delta(x_f-z)\omega_j\Big)\,.
\end{align}
Considering the limit in Eq.\eqref{limit} within Eq. \eqref{0-det}, we obtain
\begin{align}
    \mathrm{det}_{0}(\mathcal{J})=(8\pi^2)^{N_f}&\rho^{3N_f}\prod_{f=1}^{N_f}\int_{x_f}\int_z \varphi_R^\dagger\Delta(x_f-z)\rho^{-1}\mathcal{J}_{fg}(x_f)\Delta(x_g-z)\varphi_R\,,
\end{align}
where the factor $\rho^{-1}$ is included because of the dimensional difference between the quark zero mode and the quark propagator. We have also neglected the averaging over gauge rotation. This is possible because one can show that the integration over the gauge group element always contains the singlet part, which is identical to the determinant term \cite{Creutz:1978ub}.
A comparison with the previous equation implies that
\begin{align}
\sum_{j=1}^2\omega_j\overline{\omega}_j=\varphi_R\varphi^\dagger_R=\mathbb{P}_R\,.
\end{align}
The generating function in the semi-classical limit (for $N_f=3$) reads
\begin{align}\nonumber
   \mathcal{Z}[\mathcal{J}]=\int d\rho \,n(\rho)\mathrm{det}_{0}(\mathcal{J})=k_0\prod_{f=1}^{3} \int d^4z\int d^4x_f\overline{\omega}\Delta(x_f-z)\rho^{-1}\Delta(x_f-z)\omega\,,
\end{align}
where $k_0$ is defined in Eq. (\ref{eq:kJ}). 
The numerical value 7.07534 in Eq. \eqref{inst-den} is taken from Ref. \cite{Boccaletti:2020mxu} (upon settings $N_f = N_c =3$ within the modified subtraction scheme. Moreover, the running coupling at the two-loop order for small $x$-values is:
\begin{align}
   g^2(x)=\frac{(4\pi)^2}{9\log(x^{-2})}\Big(1-\frac{64}{81}\frac{\log(\log(x^{-2}))}{\log(x^{-2})}\Big)\,.
   \label{running_coupling}
\end{align}
One can proceed similarly for higher spins and estimate various couplings within the DGI model. For instance, the Lagrangian that includes spin-1 heterochiral nonet in terms of quark zero-modes reads:
\begin{align}\nonumber  \mathcal{L}^{J=1}_{\text{eff}}
&=-k_1\Big(\Psi^\dagger(x_1,z)\,\omega \overline{\omega}\,\Psi(x_1,z)\Big) 
 \Big(\Psi^\dagger(x_2,z)\,\omega \overleftrightarrow{D}_{\mu}(x_2) \overline{\omega}\,\Psi(x_2,z)\Big) \Big(\Psi^\dagger(x_3,z)\,\omega \overleftrightarrow{D}^{\mu}(x_3) \overline{\omega}\,\Psi(x_3,z)\Big)=\\
&\qquad\qquad=-\frac{k_1}{3!}\Big(\epsilon\Big[\big(\overline{q}_L q_R\big)\big(\overline{q}_L\overleftrightarrow{D}_\mu q_R\big)\big(\overline{q}_L \overleftrightarrow{D}^{\mu} q_R\big)\Big]+ R\leftrightarrow L \Big)
\,,
    \end{align}
where the coupling $k_1$  is defined in Eq. (\ref{eq:kJ}).
In the last line, the property of the Fermi-statistics is used for the quark fields. 
An analogous expression for the case of spin-2 holds:
\begin{align}  \nonumber
\mathcal{L}^{J=2}_{\text{eff}}
  =-&k_2\Big(\Psi^\dagger(x_1,z)\,\omega \overline{\omega}\,\Psi(x_1,z)\Big)
      \Big(\Psi^\dagger(x_2,z)\,\omega  (\overleftrightarrow{D}_{\mu}\overleftrightarrow{D}_{\nu}-g_{\mu\nu}\overleftrightarrow{D}^2/4) (x_2) \overline{\omega}\,\Psi(x_2,z)\Big)\\\nonumber    
    &  \Big(\Psi^\dagger(x_3,z)\,\omega  (\overleftrightarrow{D}^{\mu}\overleftrightarrow{D}^{\nu}-g_{\mu\nu}\overleftrightarrow{D}^2/4)(x_3)  \overline{\omega}\,\Psi(x_3,z)\Big)
    =\\
& 
 = -\frac{k_2}{3!} \Big(\epsilon \Big[\big(\overline{q}_L  q_R\big)
\big(\overline{q}_L(\overleftrightarrow{D}_{\mu}\overleftrightarrow{D}_{\nu}-g_{\mu\nu}\overleftrightarrow{D}^2/4)q_R\big) \big(\overline{q}_L(\overleftrightarrow{D}_{\mu}\overleftrightarrow{D}_{\nu}-g_{\mu\nu}\overleftrightarrow{D}^2/4)q_R\big)\Big] +R\leftrightarrow L \Big)
    \,.
    \end{align}

The interaction Lagrangian for the pseudoscalar glueball in Ref. (\ref{eq:gl}) is, in terms of the quark and gluonic fields:
\begin{eqnarray}\nonumber  \mathcal{L}^{J=0}_{\text{glu}}
&=& \frac{-ik_0^\prime}{4} {\rm tr} (G_{\mu\nu}\Tilde{G}^{\mu\nu}) \; \Big(\Psi^\dagger(x_1,z)\,\omega \overline{\omega}\,\Psi(x_1,z)\Big) 
 \Big(\Psi^\dagger(x_2,z)\,\omega  \overline{\omega}\,\Psi(x_2,z)\Big) \Big(\Psi^\dagger(x_3,z)\,\omega  \overline{\omega}\,\Psi(x_3,z)\Big)\\
 &=& -i\frac{k_0^\prime}{4\cdot3!}{\rm tr} (G_{\mu\nu}\Tilde{G}^{\mu\nu})\Big(\epsilon\Big[\big(\overline{q}_L q_R\big)\big(\overline{q}_L q_R\big)\big(\overline{q}_L q_R\big)\Big]+ R\leftrightarrow L \Big)
\, .
    \end{eqnarray}
The pseudoscalar
glueball is linked to the topological charge density via:
\begin{align}
    \Tilde{G}_0=\frac{1}{4\lambda_g}
    \; {\rm tr} (G_{\mu\nu}\Tilde{G}^{\mu\nu}) \, ,
\end{align}    
    where $\lambda_{g}=\bra{0}g^2
    {\rm tr} (G_{\mu\nu}\Tilde{G}^{\mu\nu})\ket{0^{-+}} $ is expected to be larger \cite{Zetocha:2002as} than the scalar glueball coupling strength $15\,  \text{GeV}^3$ calculated within the DGI model in Ref. \cite{Schafer:1994fd}. Considering the modification due to the gluonic fields in Eq. \eqref{eq:kf},
    \begin{align}   
    k_0^\prime= (8\pi^2)^{3}\int_0^{\Lambda^{-1}_{\overline{\text{MS}}}}d\rho\; n(\rho)\;\frac{\pi^4 \rho^{13 }}{8g^4(\rho)}\,,
\end{align}
we obtain the following numerical value for the decay constant of the pseudoscalar glueball for a DGI:
\begin{equation}
    c_{g}=\frac{k_0^\prime M_0^6\lambda_{g}}{48}\approx 11 \,.
    \label{a1}
\end{equation}

    \section{Extended forms of the Lagrangians}
    In this section, we present the extended forms of the Lagrangians used in the main part of the paper. The interaction with the scalar mesons has been disregarded, as their precise nonet identification  is not yet known.

The extended form of  Eq. \eqref{anom_j1} describes the anomalous decay of the orbitally excited vector mesons:
\begin{align}\nonumber
    \mathcal{L}^{J=1}_{\text{eff}}&=-a_1 \Big(\epsilon\Big[\Phi\Phi_{\mu} \Phi^{\mu}\Big]+\text{c.c.}\Big) =\\\nonumber&
    =-\frac{a_1}{6}\Big\{ \rho_{1E}^{0} \left( \Bar{K}_{1B}^0\,K^{0}+K_{1B}^0\,\Bar{K}^0-K_{1B}^{+}\,K^{-} -K_{1B}^{-}\,K^{+} +\sqrt{2}\,h_{1s}\,\pi^{0} +\sqrt{2}\, b_{1}^{0}\, \eta^\prime\cos{\beta_0} +\sqrt{2}\,b_{1}^{0}\, \eta\sin{\beta_0}  \right)+\\\nonumber&
\qquad \rho_{1E}^{+} \left(-\sqrt{2}\, K_{1B}^{-}\,K^{0}-\sqrt{2}\, K_{1B}^{0} \,K^{-}+\sqrt{2}\,h_{1s}\, \pi^{-}+\sqrt{2}\,
b_{1}^{-}\,\eta^\prime\cos{\beta_0} +\sqrt{2}\,b_{1}^{-}\, \eta\sin{\beta_0}  \right)+\\\nonumber&
\qquad \rho_{1E}^{-} \left(-\sqrt{2}\, K_{1B}^{+} \,\Bar{K}^0-\sqrt{2}\, \Bar{K}_{1B}^{0}\,K^{+}+\sqrt{2}\,h_{1s}\,\pi^{+} +\sqrt{2}\,b_{1}^{+}\, \eta^\prime \cos{\beta_0}+\sqrt{2}\, b_{1}^{+} \,\eta\sin{\beta_0}  \right)+\\\nonumber&\qquad
\omega_{1sE} \left(\sqrt{2}\,
b_1^0\,\pi^0+\sqrt{2}\, b_1^{+}\,\pi^{-}+\sqrt{2}\,b_{1}^{-}\,\pi^{+}-\sqrt{2}\,h_{1n}\, \eta \cos{\beta_0}  +\sqrt{2}\, h_{1n}\, \eta^\prime\sin{\beta_0} \right)+\\\nonumber&
\qquad \omega_{1nE} \Big( \Bar{K}_{1B}^0 \,K^0+K_{1B}^{0}\,
\Bar{K}^0+K_{1B}^{+}\,K^{-} +K_{1B}^{-}\,K^{+} -\\\nonumber
&\qquad\sqrt{2}\, h_{1s}\,\eta \cos{\beta_0} -\sqrt{2}\, h_{1n}\,\eta^\prime \cos{\beta_0}
-\sqrt{2}\,h_{1n}\, \eta \sin{\beta_0}  +\sqrt{2}\, h_{1s}\,\eta^\prime\sin{\beta_0} \Big)+\\\nonumber&\qquad
 \Bar{K}_{1E}^0 \left(b_{1}^{0} \,K^0+h_{1n}\,K^0 -\sqrt{2}\, b_{1}^{-}\,K^+ +K_{1B}^{0} \,\pi^0-\sqrt{2}\,
K_{1B}^{+} \,\pi^{-}+K_{1B}^{0}\, \eta\cos{\beta_0} -K_{1B}^{0}\, \eta^\prime  \sin{\beta_0}\right)+\\\nonumber&\qquad
K_{1E}^0 \left(b_1^{0}\, \Bar{K}^0+h_{1n} \,\Bar{K}^0-\sqrt{2}\, b_1^{+}\,K^{-} +\Bar{K}_{1B}^0\,\pi^{0}-\sqrt{2}\, K_{1B}^{-}\,
\pi^{+}+\Bar{K}_{1B}^0 \,\eta  \cos{\beta_0}-\Bar{K}_{1B}^{0}\, \eta^\prime  \sin{\beta_0}\right)+\\\nonumber&\qquad
K_{1E}^{+} \left(-\sqrt{2}\,b_{1}^{-}\, \Bar{K}^0-b_{1}^{0}\,K^{-} +h_{1n}\,K^{-}-K_{1B}^{-} \,\pi^{0}-\sqrt{2}\,\Bar{K}_{1B}^0 \,
\pi^{-}+K_{1B}^{-}\, \eta  \cos{\beta_0}-K_{1B}^{-}\, \eta^\prime  \sin{\beta_0}\right)+\\&
\qquad K_{1E}^{-}
\left(-\sqrt{2}\, b_{1}^{+}\,K^{0}-b_{1}^{0} \,K^{+}+h_{1n}\,K^{+}-K_{1B}^{+} \,\pi^{0}-\sqrt{2}\, K_{1B}^{0}\, \pi^{+}+K_{1B}^{+}\,
\eta  \cos{\beta_0}-K_{1B}^{+}\, \eta^\prime  \sin{\beta_0}\right)\Big\}\,.
\end{align}

An analogous Lagrangian for spin-2 fields in Eq. \eqref{anom_j2} is, explicitly:
  \begin{align}\nonumber
       \mathcal{L}^{J=2}_{\text{eff}}&= -a_2 \,\Big(\epsilon\Big[\Phi \Phi_{\mu\nu} \Phi^{\mu\nu}\Big]+\text{c.c.}\Big)=\\\nonumber
       &=-\frac{a_2}{6}\Big\{a_2^{0}  \left(\Bar{K}_{2B}^0\,K^0 +K_{2B}^0\,\Bar{K}^0-K_{2B}^{+}\,K^{-}-K^{-}_{2B}\,K^{+}+\sqrt{2}\, \eta_{2s}\,\pi^0+\sqrt{2}\, \pi^0_2\,\eta^\prime \cos{\beta_0} +\sqrt{2}\, \pi_{2}^0 \,\eta\sin{\beta_0}\right)+\\\nonumber
& \qquad a_2^{+} \left(-\sqrt{2}\, (K_{2B}^{-}\,K^0 + K_{2B}^0\,K^{-} -\eta_{2s}\,\pi^{-}-
\pi_2^{-}\,\eta^\prime\cos{\beta_0} -  \pi_2^{-}\,\eta  \sin{\beta_0})\right)+\\\nonumber
&\qquad a_2^{-} \left(-\sqrt{2}\,
(K_{2B}^+\,\Bar{K}^0 +\Bar{K}_{2B}^0\, K^{+}- \eta_{2s}\,\pi^{+} -\pi_{2}^+\,\eta^\prime \cos{\beta_0}- \pi_2^{+}\,\eta  \sin{\beta_0})\right)
       \\\nonumber
      & \qquad \Bar{K}_2^0\left(\eta_{2n}\,K^0+K_{2B}^0\,\pi^0 +\pi_2^{0}\,K^0 -\sqrt{2}\, \pi_{2}^{-}\,K^{+}-\sqrt{2}\, K_{2B}^{+}\,\pi^{-}+K_{2B}^{0}\, \eta  \cos{\beta_0}-K_{2B}^0\eta^\prime \sin{\beta_0}\right)+\\\nonumber
&\qquad
K_2^0 \left( \eta_{2n}\,\Bar{K}^{0}+\Bar{K}_{2B}^0\,\pi^0+\pi_{2}^0\,\Bar{K}^0-\sqrt{2}\,\pi_{2}^{+}\,K^{-}  -\sqrt{2}\,
K_{2B}^{-}\,\pi^{+} + \Bar{K}_{2B}^{0}\,\eta  \cos{\beta_0}-\Bar{K}_{2B}^0\,\eta^{\prime}\sin{\beta_0}\right)+\\\nonumber
& \qquad K_2^{+}
\left(\eta_{2n}\,K^{-} -K_{2B}^{-}\,\pi^{0} -\pi_{2}^0\,K^{-}-\sqrt{2}\,\pi_{2}^{-}\, \Bar{K}^0-\sqrt{2}\, \Bar{K}_{2B}^{0}\,\pi^{-}
+K_{2B}^{-}\,\eta \cos{\beta_0}-K_{2B}^{-}\,\eta^\prime\sin{\beta_0}\right)+\\\nonumber
&\qquad K_2^{-}
\left(\eta_{2n}\,K^{+} -K_{2B}^{+}\,\pi^{0} -\pi_{2}^{0}\,K^{+}-\sqrt{2}\,\pi_2^+\, K^{0} -\sqrt{2}\, K_{2B}^0\,\pi^{+}+K_{2B}^{+}\,\eta  \cos{\beta_0}-K_{2B}^{+}\,\eta^\prime\sin{\beta_0}\right)+\\\nonumber
& \qquad f_{2s} \left(\sqrt{2}\,(
\pi_2^0\,\pi^0 +\pi_2^{+}\,\pi^{-} + \pi_2^{-}\,\pi^{+} -  \eta_{2n}\, \eta\cos{\beta_0}+ \eta_{2n}\,\eta^{\prime} \sin{\beta_0})\right)+\\\nonumber
& \qquad f_{2n} \Big(\Bar{K}_{2B}^0\, K^0 + K_{2B}^0\,\Bar{K}^0
+K_{2B}^{+}\,K^{-}+K_{2B}^{-}\,K^{+}-\sqrt{2}\, \eta_{2s}\,\eta \cos{\beta_0}-\\
&\qquad\qquad-\sqrt{2}\, \eta_{2n} \,\eta^\prime
\cos{\beta_0}-\sqrt{2}\,\eta_{2n} \, \eta \sin{\beta_0}+\sqrt{2}\,\eta_{2s} \, \eta^\prime\sin{\beta_0}\Big)\Big\}\,.
  \end{align} 

The Lagrangian in Eq. \eqref{anom_j012} is, in detail:
  \begin{align}\nonumber
      & \mathcal{L}_{b_2}= -b_2\Big( \epsilon\Big[ \partial_\mu\big(\Phi\big) \Phi_{ \nu} \Phi^{\mu\nu}\Big]+\text{c.c.}\Big)=\\\nonumber
      & \qquad = -\frac{b_2}{12}\Big\{a_2^{0} \left(\Bar{K}_{1B}^0\, K^0+K_{1B}^0\,\Bar{K}^0 -K_{1B}^{+}\,K^{-} -K_{1B}^{-}\,K^{+}+\sqrt{2}\,h_{1s}\,\pi^0
+\sqrt{2}\,b_{1}^{0}\,\eta^\prime \cos{\beta_0}+\sqrt{2}\,b_{1}^{0}\, \eta  \sin{\beta_0}\right)+\\\nonumber
&\qquad\qquad a_{2}^{+}
 \left(-\sqrt{2}\, (K_{1B}^{-} \,K^0+K_{1B}^0\,K^{-}-h_{1s}\,\pi^{-} - b_1^{-}\,\eta^\prime \cos{\beta_0} - b_1^{-}\, \eta\sin{\beta_0} ) \right)+\\\nonumber
& \qquad\qquad a_2^{-} \left(-\sqrt{2}\, (K_{1B}^{+}\,\Bar{K}^0 + \Bar{K}_{1B}^0\,K^{+}
- h_{1s}\,\pi^{+}-b_{1}^{+}\, \eta^\prime\cos{\beta_0} - b_1^{+} \,\eta\sin{\beta_0} ) \right)+\\\nonumber
& \qquad\qquad f_{2s} \left(\sqrt{2}\,( b_1^{0}\,\pi^{0} +b_1^{+}\,\pi^{-} + b_1^{-}\,\pi^{+} -h_{1n}\,
\eta\cos{\beta_0}  + h_{1n}\,\eta^\prime\sin{\beta_0}) \right)+\\\nonumber
& \quad\quad\qquad f_{2n}
\Big(\Bar{K}_{1B}^0\,K^0+K_{1B}^0\,\Bar{K}^0+K_{1B}^{+}\,K^{-}+K_{1B}^{-}\,K^{+}-\sqrt{2}\,h_{1s} \,\eta\cos{\beta_0} -\sqrt{2}\, h_{1n} \,\eta^\prime\cos{\beta_0}\\\nonumber
&\qquad\quad\qquad-\sqrt{2}\, h_{1n}\, \eta  \sin{\beta_0}+\sqrt{2}\, h_{1s}\eta^\prime\sin{\beta_0}\Big)+\\\nonumber
&\qquad\qquad \Bar{K}_{2B}^{0} \left(K_{1E}^{0}\,\pi^{0} -\sqrt{2}\, K_{1E}^{+}\,\pi^{-}+\rho_{1E}^0\,K^{0}-\sqrt{2}\,\rho_{1E}^{-}\, K^{+}+ \omega_{1nE}\,K^{0}+K_{1E}^0 \,\eta \cos{\beta_0}  -K_{1E}^0\, \eta^\prime \sin{\beta_0}\right)+\\\nonumber
&\qquad\qquad \Bar{K}_{2}^0  \left(b_1^{0}\, K^{0}+h_{1n} \,K^{0}-\sqrt{2}\, b_1^{-}\,K^{+}
+K_{1B}^{0}\,\pi^{0}-\sqrt{2}\,K_{1B}^{+}\, \pi^{-}+K_{1B}^{0}\, \eta\cos{\beta_0} -K_{1B}^{0}\,\eta^{\prime}\sin{\beta_0}
\right)+\\\nonumber
&\qquad\qquad K_{2B}^{0}  \left(\Bar{K}_{1E}^0\,\pi^0-\sqrt{2}\, K_{1E}^{-}\,\pi^{+}+\rho_{1E}^{0}\,\Bar{K}^{0}-\sqrt{2}\,\rho_{1E}^{+} \,K^{-}+\omega_{1nE}\,\Bar{K}^{0}+\Bar{K}_{1E}^0\, \eta  \cos{\beta_0}-\Bar{K}_{1E}^{0}\,\eta^{\prime}
\sin{\beta_0}\right)+\\\nonumber
&\qquad\qquad  K_{2}^{0}\left(b_1^0 \,\Bar{K}^0+h_{1N}\,\Bar{K}^{0} -\sqrt{2}\, b_{1}^{+}\,K^{-}
+\Bar{K}_{1B}^0\,\pi^0-\sqrt{2}\,K_{1B}^{-}\,\pi^{+} +\Bar{K}_{1B}^0 \,\eta  \cos{\beta_0}-\Bar{K}_{1B}^0\, \eta^\prime \sin{\beta_0}\right)+\\\nonumber
&\qquad\qquad K_{2}^{+} \left(-\sqrt{2}\, b_1^-\,\Bar{K}^0 -b_1^0\,K^{-} +h_{1N}\,K^- -K_{1B}^{-}\,\pi^0
-\sqrt{2}\, \Bar{K}_{1B}^0\, \pi^{-} +K_{1B}^{-} \,\eta\cos{\beta_0} -K_{1B}^{-} \,\eta^\prime\sin{\beta_0} \right)+\\\nonumber
&\qquad\qquad  K_2^{-} \left(-\sqrt{2}\, b_1^+\,K^0 -b_1^0 \,K^{+}+h_{1n}\,K^+-K_{1B}^+\,\pi^0-\sqrt{2}\,
K_{1B}^0\,\pi^++K_{1B}^+\, \eta\cos{\beta_0}  -K_{1B}^+\,\eta^\prime \sin{\beta_0} \right)+\\\nonumber
&\qquad\qquad K_{2B}^{+}
 \left(-K^{-}_{1E}\,\pi^0-\sqrt{2}\, \Bar{K}_{1E}^0\,\pi^{-} - \rho_{1E}^{0}\,K^{-}-\sqrt{2}\,\rho_{1E}^{-}\,\Bar{K}^0  +\omega_{1nE}\,K^{-}+K_{1E}^{-}\, \eta  \cos{\beta_0}-K_{1E}^{-}\,\eta^\prime\sin{\beta_0} \right)+\\\nonumber
&\qquad\qquad K_{2B}^{-}
 \left(-K_{1E}^{+}\,\pi^0 -\sqrt{2}\, K_{1E}^0\,\pi^{+}-\rho_{1E}^{0}\,K^{+} -\sqrt{2}\,\rho_{1E}^{+}\, K^{0}+
\omega_{1nE}\,K^{+}+K^{+}_{1E}\, \eta\cos{\beta_0}  -K_{1E}^{+}\,\eta^\prime\sin{\beta_0} \right)+\\\nonumber
& \qquad\qquad\pi^{0}_2
 \left(\Bar{K}^0_{1E}\,K^{0} +K_{1E}^{0}\,\Bar{K}^0-K_{1E}^{+}\,K^{-}-K_{1E}^{-}\,K^{+}+\sqrt{2}\,\omega_{1sE}\,\pi^0 +\sqrt{2}\,  \rho_{1E}^0\,\eta^\prime \cos{\beta_0} +\sqrt{2}\, \rho_{1E}^{0}\,\eta \sin{\beta_0} \right)+\\\nonumber
& \qquad\qquad
\pi^{+}_2 \left(-\sqrt{2}\,(  K_{1E}^{-}\,K^0 + K_{1E}^{0}\,K^{-} -  \omega_{1sE}\,\pi^{-}-\rho_{1E}^{-} \,\eta^\prime \cos{\beta_0} -\rho_{1E}^{-}\,\eta  \sin{\beta_0})\right)+\\\nonumber
&\qquad\qquad
\pi^{-}_2
 \left(-\sqrt{2}\,( K_{1E}^{+}\,\Bar{K}^0 +\Bar{K}_{1E}^0\,K^{+} - \omega_{1sE}\,\pi^{+}-\rho^{+}_{1E}\,\eta^\prime \cos{\beta_0}-  \rho^{+}_{1E}\,\eta  \sin{\beta_0})\right)+\\\nonumber
&\qquad\qquad \eta_{2s} \left(\sqrt{2}\,( \rho_{1E}^{0}\,\pi^0 + \rho_{1E}^{-}\,\pi^{+}  +\rho_{1E}^{+}\,\pi^{-} -
\omega_{1nE}\,\eta \cos{\beta_0} +\omega_{1nE}\,\eta^\prime\sin{\beta_0}) \right)+\\\nonumber
&
\qquad\qquad\eta_{2n}\Big( \Bar{K}_{1E}^0\,K^{0}+K_{1E}^0\,\Bar{K}^0 +K_{1E}^{+}\,K^{-}+K_{1E}^{-}\,K^{+} -\sqrt{2}\,
\omega_{1nE}\,\eta^\prime\cos{\beta_0} -\\
&\qquad\qquad\qquad-\sqrt{2}\,\omega_{1sE}\,\eta\cos{\beta_0} -\sqrt{2}\, \omega_{1nE}\,\eta  \sin{\beta_0}
+\sqrt{2}\,  \omega_{1sE}\,\eta^\prime\sin{\beta_0} \Big)\Big\}.  
  \end{align}

The decay of a spin-2 heterochiral meson into two pseudoscalars
is given by Eq. \eqref{anom_j022}.
In terms of the component fields, this equals:  
     \begin{align}\nonumber
    \mathcal{L}_{c_2}&= - c_2\Big(\epsilon\Big[\partial_\mu\big(\Phi\big)\partial_\nu\big(\Phi\big)\Phi^{\mu\nu}\Big]+\text{c.c.}\Big)=\\\nonumber
&=-\frac{c_2}{24}\Big\{ a_2^{0\,\mu\nu} \left(4 (\partial_\mu K^0\, \partial_\nu
\Bar{K}^0-\partial_\mu K^{-}\,\partial_\nu K^{+} )+4 \sqrt{2}\, \,\partial_\mu\eta^\prime\, \partial_\nu \pi^0\cos{\beta_0}+4 \sqrt{2}\, \, \partial_\mu\eta\, \partial_\nu\pi^{0}\sin{\beta_0}]\right)+\\\nonumber
&\qquad a_2^{+\,\mu\nu} \left(-4 \sqrt{2}\, \, \partial_\mu K^{0}\, \partial_\nu K^{-} +4 \sqrt{2}\,\, \partial_\mu\eta^{\prime} \, \partial_\nu\pi^{-}\cos{\beta_0}+4
\sqrt{2}\,\, \partial_\mu\eta \,\partial_\nu\pi^{-} \sin{\beta_0}\right)+\\\nonumber
&\qquad a_2^{-\,\mu\nu} \left(-4 \sqrt{2}\,\, \partial_\mu\Bar{K}^0\, \partial_\nu K^{+} +4 \sqrt{2}\,\, \partial_\mu \eta^\prime \, \partial_\nu\pi^{+} \cos{\beta_0}+4 \sqrt{2}\,\, \partial_\mu \eta  \, \partial_\nu\pi^{+}\sin{\beta_0}\right)+\\\nonumber
&\qquad \Bar{K}_2^{0\,\mu\nu} \Big(4 \, \partial_\mu K^0 \, \partial_\nu\pi^0 -4 \sqrt{2}\,\, \partial_\mu K^{+}\, \partial_\nu\pi^{-} +4 \, \partial_\mu K^0\, \partial_\nu\eta  \cos{\beta_0}-4\, \partial_\mu K^0\, \partial_\nu\eta^\prime\sin{\beta_0}\Big)+\\\nonumber
&\qquad K_2^{0\,\mu\nu} \left(4 \, \partial_\mu\Bar{K}^0\, \partial_\nu\pi^0-4 \sqrt{2}\,\, \partial_\mu K^{-}\, \partial_\nu\pi^{+}
+4\, \partial_\mu\Bar{K}^0\, \partial_\nu\eta \cos{\beta_0} -4 \, \partial_\mu\Bar{K}^0\, \partial_\nu\eta^\prime\sin{\beta_0}\right)+\\\nonumber
&\qquad K_2^{+\,\mu\nu}
 \left(-4\, \partial_\mu K^{-} \, \partial_\nu\pi^0-4 \sqrt{2}\, \, \partial_\mu\Bar{K}^0\, \partial_\nu\pi^{-}+4\, \partial_\mu K^{-\,\mu\nu}\, \partial_\nu \eta \cos{\beta _p}-4\, \partial_\mu K^{-} \, \partial_\nu\eta\prime\sin{\beta_0}\right)+\\\nonumber
&\qquad K_2^{-\,\mu\nu} \left(-4\, \partial_\mu K^{+} \, \partial_\nu\pi^{0}-4 \sqrt{2}\,\, \partial_\mu K^{0} \, \partial_\nu\pi^{+}+4\, \partial_\mu K^{+}
\, \partial_\nu\eta \cos{\beta_0}-4\, \partial_\mu K^{+}\, \partial_\nu \eta^\prime \sin{\beta_0}\right)+\\\nonumber
&\qquad f_{2s}^{\,\mu\nu} \Big(-\sqrt{2}\,
\left(1+\cos{2\beta_0}\right)\, \partial_\mu\eta\, \partial_\nu\eta-\sqrt{2}\,
\left(1-\cos{2\beta_0}\right)\, \partial_\mu\eta^\prime\, \partial_\nu\eta^\prime+2 \sqrt{2}\, \, \partial_\mu\pi^{0}\, \partial_\nu\pi^0
\\\nonumber
&\qquad\qquad+4 \sqrt{2}\,\, \partial_\mu \pi^{-}\, \partial_\nu\pi^{+} +2 \sqrt{2}\, \, \partial_\mu\eta\, \partial_\nu\eta^\prime\sin{2 \beta_0}\Big)+\\
&\qquad f_{2n}^{\,\mu\nu}
\left(4 (\partial_\mu K^0 \partial_\nu\Bar{K}^0 +\partial_\mu K^{-}\partial_\nu K^{+})-4 \sqrt{2}\, \partial_\mu\eta  \partial_\nu\eta^\prime \cos{2 \beta_0}+2 \sqrt{2}\, \left(-\partial_\mu\eta\partial_\nu\eta +\partial_\mu\eta^\prime\partial_\nu\eta^\prime\right) \sin{2 \beta_0}\right)\Big\}\,.
    \end{align}
    
The anomalous interactions
of Eq. \eqref{eq:gl} equal:
\begin{align}
\mathcal{L}_{c_g}=&-\mathrm{i}c_{g}\Tilde{G}_0\Big(\text{det}\,\Phi-\text{det}\,\Phi^\dagger \Big)=-\frac{\mathrm{i}c_{g}}{2\sqrt{2}}%
\tilde{G}_0\Big(\sqrt{2}Z_{K_{S}}Z_{K}a_{0}^{0}K_{S}^{0}\overline{K}^{0}+\sqrt
{2}Z_{K}Z_{K_{S}}a_{0}^{0}K^{0}\overline{K}_{S}^{0}-2Z_{K_{S}}Z_{K}a_{0}%
^{+}K_{S}^{0}K^{-}\nonumber\\
&  -2Z_{K_{S}}Z_{K}a_{0}^{+}K_{S}^{-}K^{0}-2Z_{K_{S}}Z_{K}a_{0}^{-}%
\overline{K}_{S}^{0}K^{+}-\sqrt{2}Z_{K_{S}}Z_{K}a_{0}^{0}K_{S}^{-}K^{+}%
-\sqrt{2}Z_{K}^{2}Z_{\eta_{N}}K^{0}\overline{K}^{0}\eta_{N}\nonumber\\
&  +\sqrt{2}Z_{K_{S}}^{2}Z_{\eta_{N}}K_{S}^{0}\overline{K}_{S}^{0}\eta
_{N}-\sqrt{2}Z_{K}^{2}Z_{\eta_{N}}K^{-}K^{+}\eta_{N}+Z_{\eta_{S}}{a_{0}^{0}%
}^{2}\eta_{S}+2Z_{\eta_{S}}a_{0}^{-}a_{0}^{+}\eta_{S}\nonumber\\
&  +Z_{\eta_{N}}^{2}Z_{\eta_{S}}\eta_{N}^{2}\eta_{S}-\sqrt{2}Z_{K}^{2}Z_{\pi
}K^{0}\overline{K}^{0}\pi^{0}+\sqrt{2}Z_{K_{S}}^{2}Z_{\pi}K_{S}^{0}%
\overline{K}_{S}^{0}\pi^{0}+\sqrt{2}Z_{K}^{2}Z_{\pi}K^{-}K^{+}\pi
^{0}\nonumber\\
&  -Z_{\eta_{S}}Z_{\pi}^{2}\eta_{S}\,{\pi^{0}}^{2}+2Z_{K}^{2}Z_{\pi}%
\overline{K}^{0}K^{+}\pi^{-}+2Z_{K}^{2}Z_{\pi}K^{0}K^{-}\pi^{+}-2Z_{K_{S}}%
^{2}Z_{\pi}K_{S}^{0}K_{S}^{-}\pi^{+}\nonumber\\
&  -2Z_{\eta_{S}}Z_{\pi}^{2}\eta_{S}\pi^{-}\pi^{+}-2Z_{K_{S}}Z_{K}a_{0}%
^{-}K_{S}^{+}\overline{K}^{0}+\sqrt{2}Z_{K_{S}}^{2}Z_{\eta_{N}}K_{S}^{+}%
K_{S}^{-}\eta_{N}-\sqrt{2}Z_{K_{S}}^{2}Z_{\pi}K_{S}^{+}K_{S}^{-}\pi
^{0}\nonumber\\
&  -2Z_{K_{S}}^{2}Z_{\pi}K_{S}^{+}\overline{K}_{S}^{0}\pi^{-}-\sqrt{2}%
Z_{K_{S}}Z_{K}a_{0}^{0}K_{S}^{+}K^{-}+\sqrt{2}Z_{K}Z_{K_{S}}K^{-}K_{S}^{+}%
\phi_{N}+\sqrt{2}Z_{K}Z_{K_{S}}K^{-}K_{S}^{+}\sigma_{N}\nonumber\\
&  +\sqrt{2}Z_{K_{S}}Z_{K}K_{S}^{0}\overline{K}^{0}\phi_{N}+\sqrt{2}Z_{K_{S}%
}Z_{K}K_{S}^{0}\overline{K}^{0}\sigma_{N}+\sqrt{2}Z_{K}Z_{K_{S}}K^{0}%
\overline{K}_{S}^{0}\phi_{N}+\sqrt{2}Z_{K}Z_{K_{S}}K^{0}\overline{K}_{S}%
^{0}\sigma_{N}\nonumber\\
&  +\sqrt{2}Z_{K_{S}}Z_{K}K_{S}^{-}K^{+}\phi_{N}+\sqrt{2}Z_{K_{S}}Z_{K}%
K_{S}^{-}K^{+}\sigma_{N}-Z_{\eta_{S}}\eta_{S}\phi_{N}^{2}-Z_{\eta_{S}}\eta
_{S}\sigma_{N}^{2}-2Z_{\eta_{S}}\eta_{S}\phi_{N}\sigma_{N}\nonumber\\
&  +2Z_{\pi}a_{0}^{0}\pi^{0}\phi_{S}+2Z_{\pi}a_{0}^{0}\pi^{0}\sigma
_{S}+2Z_{\pi}a_{0}^{+}\pi^{-}\phi_{S}+2Z_{\pi}a_{0}^{+}\pi^{-}\sigma
_{S}+2Z_{\pi}a_{0}^{-}\pi^{+}\phi_{S}+2Z_{\pi}a_{0}^{-}\pi^{+}\sigma
_{S}\nonumber\\
&  -2Z_{\eta_{N}}\eta_{N}\phi_{N}\phi_{S}-2Z_{\eta_{N}}\eta_{N}\phi_{N}%
\sigma_{S}-2Z_{\eta_{N}}\eta_{N}\sigma_{N}\phi_{S}-2Z_{\eta_{N}}\eta_{N}%
\sigma_{N}\sigma_{S}\Big)\text{ ,}%
\end{align}
where $Z_{\pi} =Z_{\eta_N}=1.709$, $Z_{\eta_S}=1.539$ and $Z_{K}=1.604$ are from Ref. \cite{Eshraim:2012jv}.
Other interaction Lagrangians for the pseudoscalar glueball can be obtained, such as
\begin{equation}
    \Tilde{G}_0\Big( \epsilon[\Phi\, \Phi_\mu \, \Phi^\mu]
-\text{c.c.}\Big) \; .
\end{equation}

\section{Decay Formulas}
The Lagrangian in Eq. \eqref{anom_j1} gives the decay rate
\begin{align}    \Gamma^{\text{tl}}_{P_{1}\rightarrow B_1+P}=\frac{a_1^{\,2}\,\kappa_{i}}{8\pi m_{p_1}}\Big(1+\frac{|\Vec{k}(m_{p_1},m_{b_1},m_{p})|^2}{3m_{b_1}^2}\Big) \Theta \Big(m_{p_1}-m_{b_1}-m_{p}\Big)\,,
\end{align}
where the Heaviside step-function is denoted by $\Theta(x)$, $\kappa_i$'s are given in Table \ref{tab:vkappa}, and the modulus of the outgoing particle momentum is
\begin{equation}
|\vec{k}(m_{a},m_{b},m_{c})|:=\frac{\Big((m_{a}^{2}-m_{b}^{2}-m_{c}^{2}%
)^{2}-4m_{b}^{2}m_{c}^{2}\Big)^\frac{1}{2}}{2\,m_{a}}\ \text{ .} \label{eq:kf}%
\end{equation}
\begin{table}[h]		
		\renewcommand{\arraystretch}{2.5}
  \centering
				 \begin{tabular}{|c|c|c|c|}
			\hline decay process 				& $\kappa_{i}$ 			
					&		
				 decay process 				& $\kappa_{i}$ 				
				\\
				\hline
    $\rho(1700) \rightarrow h_1(1415) \, \pi$ & $\Big(\frac{\sqrt{2}\,}{3!}\Big)^2$ & $\eta_2(1870) \rightarrow \rho(1700) \, \pi$ & 3$\Big(\frac{1}{3!\sqrt{2}\,}\Big)^2$									\\ \hline
     $\phi (2170) \rightarrow b_1(1235) \, \pi$	& $3(\frac{\sqrt{2}\,}{3!})^2$ & $f_{2}(2300) \rightarrow \pi \, \pi$	& $6(\frac{1}{3!\sqrt{2}\,})^2$						\\\hline				
		$\phi (2170)\rightarrow h_1(1170) \, \eta$	& $ \Big(\frac{\sqrt{2}\,\cos{\beta_0}}{3!}\Big)^2$ & $f_{2}(2300) \rightarrow \pi_2(1670) \, \pi$	& $ 3\Big(\frac{\sqrt{2}\,}{3!}\Big)^2$					\\\hline				
   $\phi (2170) \rightarrow h_1(1170) \, \eta^\prime(958)$	& $\Big(\frac{\sqrt{2}\,\sin{\beta_0}}{3!}\Big)^2$	&  $f_{2}(2300) \rightarrow \eta_2(1645) \, \eta$	& $\Big(\frac{\sqrt{2}\,\cos{\beta_0}}{3!}\Big)^2$					\\\hline
			\end{tabular}
		\caption{\label{tab:vkappa} %
			 Decay coefficients for heterochiral mesons.  		}
	\end{table}

The decay rate from the Lagrangian in Eq. \eqref{anom_j012} is   
\begin{align}
\Gamma_{\eta_2(1870)\rightarrow \rho(1700)\pi}^{\text{tl}}=\frac{ b_{2}^2\kappa_{i}\,|\vec{k}(m_{\eta_2(1870)},m_{\rho(1700)},m_{\pi})|^{3}}{120\,\pi\,m_{\eta_2(1870)}^{2}}
\Big(5+\frac{2|\vec{k}(m_{\eta_2(1870)},m_{\rho(1700)},m_{\pi})|^{2}}{m_{\rho(1700)}^2}\Big)\,.
\label{eq:dec-tpp1}%
\end{align}
For $J=2$ mesons, while experimentally
the situation is unclear for the excited tensor mesons \cite{Vereijken:2023jor}, we assume it to be $f_{2s}\equiv f_2(2300)$, and the expression of its decay into two pions reads
\begin{equation}
\Gamma^{\text{tl}}_{f_2(2300)\rightarrow \pi \pi}=\frac{ c_{2}^2\kappa_{i}\,|\vec{k}(m_{f_2(2300)},m_{\pi},m_{\pi})|^{5}}{60\,\pi\,m_{f_2(2300)}^{2}}\,
. 
\label{eq:dec-tpp}%
\end{equation}

From Eq. \eqref{anom_j2}, we get the following expression for the decay rate 
\begin{align}
\Gamma_{f_{2}(2300)\rightarrow P_2+P}^{\text{tl}}=\frac{ a_2^2\,\kappa_{i}|\vec{k}(m_{f_2(2300)},m_{p_2},m_{p})%
|}{8\,m_{f_{2}(2300)}^2\pi}\Big(1+\frac{4|\vec{k}(m_{f_2(2300)},m_{p_2},m_{p})%
|^4}{45m_{p_2}^4}
+\frac{2|\vec{k}(m_{f_2(2300)},m_{p_2},m_{p})%
|^2}{3m_{p_2}^2}\Big)
\, .
\end{align}

Decays for heterochiral mesons with spin-$1$ are given
in Table \ref{tab:vpp}, and those
with spin-$2$ mesons in Table \ref{tab:tensor}. 

 \begin{table}[h]		
		\renewcommand{\arraystretch}{2.}
  \centering
				 \begin{tabular}{|c|c|}
			\hline Decay process 				& 	Width (MeV) 				
	
				\\
				\hline
    $\rho(1700) \rightarrow h_1(1415) \, \pi$						&	$ 0.027 $  	\\ \hline
     $\phi (2170) \rightarrow b_1(1235) \, \pi$				&	$0.071$   \\\hline
		$\phi (2170)\rightarrow h_1(1170) \, \eta$							&	$0.012   $\\\hline
   $\phi (2170) \rightarrow h_1(1170) \, \eta^\prime(958)$							&	$0.010  $\\\hline
			\end{tabular}
		\caption{\label{tab:vpp} %
		Widths for the anomalous decays of spin-1
mesons in a DGI, assuming $M_1=M_0$. 		}
	\end{table}

 \begin{table}[h]		
\renewcommand{\arraystretch}{2.}
  \centering
				 \begin{tabular}{|c|c|}
			\hline Decay process 				 &	Width (MeV)	 				
			
				\\
				\hline
  		 $f_{2}(2300) \rightarrow \pi_2(1670) \, \pi$							&	$2\cdot 10^{-4}\; (M_2/M_0)^8 $\\\hline
    $f_{2}(2300) \rightarrow \eta_2(1645) \, \eta$	&	$2\cdot 10^{-5}   \; (M_2/M_0)^8 $\\\hline
      $\eta_2(1870) \rightarrow \rho(1700) \, \pi$						&	$1.5\cdot 10^{-6}  \; M_1^3 M_2^4/M_0^7$		\\ \hline
   $f_{2}(2300) \rightarrow \pi \, \pi$							&	$0.05 \; (M_2/M_0)^4 	 $\\\hline
			\end{tabular}
		\caption{\label{tab:tensor} %
			Widths for the anomalous decays of spin-2 mesons in a DGI.		}
	\end{table}

  \end{widetext}


\end{document}